\documentclass[useAMS,usenatbib]{mn2e}
\usepackage{amsmath,amsfonts,epsfig,natbib}
\def\be{\begin{equation}} 
\def\ee{\end{equation}} 
\def\ba{\begin{eqnarray}} 
\def\ea{\end{eqnarray}}

\def\reff@jnl#1{{\rm#1\/}}
\def\aj{\reff@jnl{AJ}}                  % Astronomical Journal
\def\araa{\reff@jnl{ARA\&A}}            % Annual Review of Astron and Astrophys
\def\apj{\reff@jnl{ApJ}}                        % Astrophysical Journal
\def\apjl{\reff@jnl{ApJ}}               % Astrophysical Journal, Letters
\def\apjs{\reff@jnl{ApJS}}              % Astrophysical Journal, Supplement
\def\ao{\reff@jnl{Appl.Optics}}         % Applied Optics
\def\apss{\reff@jnl{Ap\&SS}}            % Astrophysics and Space Science
\def\aap{\reff@jnl{A\&A}}               % Astronomy and Astrophysics
\def\aapr{\reff@jnl{A\&A~Rev.}}         % Astronomy and Astrophysics Reviews
\def\aaps{\reff@jnl{A\&AS}}             % Astronomy and Astrophysics, Supplement
\def\azh{\reff@jnl{AZh}}                        % Astronomicheskii Zhurnal
\def\baas{\reff@jnl{BAAS}}              % Bulletin of the AAS
\def\jrasc{\reff@jnl{JRASC}}            % Journal of the RAS of Canada
\def\memras{\reff@jnl{MmRAS}}           % Memoirs of the RAS
\def\mnras{\reff@jnl{MNRAS}}            % Monthly Notices of the RAS
\def\pra{\reff@jnl{Phys.Rev.A}}         % Physical Review A: General Physics
\def\prb{\reff@jnl{Phys.Rev.B}}         % Physical Review B: Solid State
\def\prc{\reff@jnl{Phys.Rev.C}}         % Physical Review C
\def\prd{\reff@jnl{Phys.Rev.D}}         % Physical Review D
\def\prl{\reff@jnl{Phys.Rev.Lett}}      % Physical Review Letters
\def\pasp{\reff@jnl{PASP}}        
\def\pasj{\reff@jnl{PASJ}}              % Publications of the ASJ
\def\qjras{\reff@jnl{QJRAS}}            % Quarterly Journal of the RAS
\def\skytel{\reff@jnl{S\&T}}            % Sky and Telescope
\def\solphys{\reff@jnl{Solar~Phys.}}    % Solar Physics
\def\sovast{\reff@jnl{Soviet~Ast.}}     % Soviet Astronomy
\def\ssr{\reff@jnl{Space~Sci.Rev.}}     % Space Science Reviews
\def\zap{\reff@jnl{ZAp}}                        % Zeitschrift fuer Astrophysik
\def\nat{\reff@jnl{Nature}}             % Nature 

\title[Sensitive submm survey of BAL quasars]
{A sensitive submillimetre survey of Broad Absorption Line quasars}
\author[Priddey et al.]
{Robert S. Priddey$^1$\thanks{R.S.Priddey@herts.ac.uk}, 
S.C. Gallagher$^2$, K.G. Isaak$^3$, R.G. Sharp$^4$,  
R.G. McMahon$^5$ \and 
H.M. Butner$^6$\\
\\
$^1${\it Centre for Astrophysics Research, University of Hertfordshire,
College Lane, Hatfield, Hertfordshire, AL10 9AB, UK}\\
$^2${\it Department of Physics \& Astronomy, UCLA, 430 Portola Plaza,
Los Angeles, CA, USA}\\
$^3${\it School of Physics \& Astronomy, 
Cardiff University, The Parade, Cardiff, CF24 3AA, UK}\\
$^4${\it Anglo-Australian Observatory, P.O. Box 296, Epping, NSW 1710,
Australia}\\
$^5${\it Institute of Astronomy, University of Cambridge, Madingley Road,
Cambridge, CB3 0HA, UK}\\
$^6${\it Joint Astronomy Centre, 660 N. A'ohoku Place, Hilo, Hawai'i,
USA}\\
\\}
\date{Accepted 2006 October 12. Received 2006 September 18; 
in original form 2006 August 9}
\pagerange{\pageref{firstpage}--\pageref{lastpage}}
\pubyear{2006}

\begin{document}

\maketitle

\label{firstpage}

\begin{abstract}
Using the SCUBA bolometer array on the JCMT,
we have carried out a submillimetre (submm) 
survey of Broad Absorption Line quasars (BALQs).
The sample has been chosen to match, in redshift and 
optical luminosity, an existing benchmark 850$\mu$m sample 
of radio-quiet quasars, allowing a direct comparison of the 
submm properties of BAL quasars relative to the parent 
radio-quiet population. We reach a submm 
limit $1\sigma\approx$1.5mJy at 850$\mu$m, allowing a more rigorous measure of
the submm properties of BAL quasars than previous studies.
Our submm photometry complements
extensive observations at other wavelengths,
in particular X-rays with {\it Chandra} and infrared with {\it Spitzer}. 

To compare the 850$\mu$m flux distribution of BALQs with that of the 
non-BAL quasar benchmark sample, we employ a suite of statistical methods, 
including survival analysis and a 
novel Bayesian derivation of the underlying flux distribution. 
Although there are no strong grounds for rejecting the null hypothesis
that BALQs on the whole have the same submm properties as
non-BAL quasars, we do find
tentative evidence 
(1--4 percent significance from a K--S test and survival analysis)
for a dependence of submm flux on the equivalent
width of the characteristic C {\sc IV} broad absorption line.
If this effect is real---
submm activity is linked to the 
absorption strength of the outflow--- it has  
implications either for the evolution of AGN and their connection
with star formation in their host galaxies, or for unification models
of AGN.

\end{abstract}


\section{INTRODUCTION}
\subsection{AGN feedback and Broad Absorption Line quasars}
The extreme luminosity and relatively low space density of 
Active Galactic Nuclei (AGN)
might be taken to imply that they constitute an unusual subset of all 
galaxies, and that the study of AGN gives us a lopsided view of 
cosmological processes. 
On the contrary, recent findings 
indicate that activity on the small scales surrounding the central
supermassive black hole (SMBH) and its accretion disk may have a 
significant bearing on the AGN's environment on galactic scales or even larger,
playing an important role in the evolution of the host galaxy
and the surrounding intergalactic medium.
The ubiquity of SMBHs within galactic nuclei suggests that a phase of 
AGN activity is the rule rather than the exception. Furthermore, 
the discovery of a correlation between the mass of the SMBH and the
velocity dispersion of stars in the bulge of the host 
(Ferrarese \& Merritt, 2000; Gebhardt et al. 2000)
hints that the evolution of each component--- SMBH and stars---
must, at some level, be linked.

Feedback from the AGN itself has been suggested as one mechanism
by which the $M-\sigma$ relation could be established
(Silk \& Rees, 1998; Fabian, 1999; Granato et al. 2004), 
wherein outflows from an AGN-driven wind
can unbind gas from the host galaxy, halting star formation and further
black hole growth.
The effect of AGN feedback (in the form of jets or of accretion disk winds) 
on the evolution of the host galaxy
has also been posited as a solution to the
mismatch between $\Lambda$CDM hierarchical galaxy formation models and 
observed galaxy luminosity functions (e.g. Benson et al. 2003).
The study of the evolution of AGN, their host galaxies and
their outflows, is therefore
of great significance to cosmology and galaxy formation.

Evidence for outflows is 
found in about 15 percent (Reichard et al., 2003) of luminous, 
optically selected quasars, 
in the form of Broad Absorption Lines (BALs). 
BALs, characteristically in the C {\sc IV} (1549\AA) line, exhibit
broad absorption troughs with high blueshifts,
pointing to an origin in 
high velocity ($\sim$0.1c), radiation-driven outflows.
Broadly speaking, three scenarios have been proposed to account for
the BAL effect in quasars:
(1) BALQs form a distinct class of AGN in their own right (disfavoured by
the observed similarity between BALQ and non-BALQ optical/UV continua and
emission lines: Weymann et al. 1991)
(2) BAL winds are a generic feature of {\it all} AGN, but an
orientation dependence restricts the visibility of lines to only 
$\sim$10--20 percent of the population (Weymann et al., 1991; Murray et al. 
1995)
(3) the BAL phenomenon 
represents a short-lived phase in the life-cycle of
every AGN, during which the active nucleus is still
enshrouded by gas and dust which
covers a large fraction of the sky 
(Hazard et al. 1984; Voit, Weymann \& Korista, 1993). 
Either of the latter two interpretations (characterised by Becker et al. 2000
as ``unification by orientation'' as opposed to ``unification by time'') 
would have
important implications for understanding the nature and evolution of AGN,
and their relationship with their host galaxies.

\subsection{Submillimetre observations of high-redshift quasars}
Observations in the submillimetre (submm) band have opened an 
important window on the formation of massive galaxies
(for a small cross-section of the work in this area, see e.g. 
Smail, Ivison \& Blain 1997; Hughes et al. 1998; 
Scott et al. 2002; Borys et al. 2003; Mortier et al. 2005).
Models of the coevolution of SMBHs and spheroids (e.g. Granato et al.
2004) predict that a substantial portion of the star formation within
the hosts of luminous AGN ought to take place in a dust-enshrouded,
submm-luminous mode; there could therefore be a direct link between
the population of high-redshift submm galaxies and AGN.
Deep {\it Chandra} X-ray observations of fields surveyed by
the Submm Common-User Bolometer Array (SCUBA) 
indicate that AGN are present in a high fraction of submm galaxies, even
when their bolometric luminosity is dominated by a starburst 
(Alexander et al. 2005), consistent with a SMBH gradually growing
within a massive, star-forming galaxy.
It could be, then, that submm galaxies
and quasars represent similar populations observed at different points
during their life cycle.
Within this scenario, it has been speculated that BALQs---
or some subclass of BALQs such as {\it low ionization BAL quasars}
(LoBALs: BALQs that exhibit absorption in lines such as Mg {\sc II} 
in addition to the usual {\sc C IV})--- 
represent a youthful transitionary phase during which the
AGN, initially embedded within a region of high star formation,
expels its obscuring shroud of gas and dust (Voit, Weymann \& Korista, 1993)
and emerges to become a classical, optically selected quasar,
as in the evolutionary scheme of Sanders et al. (1988).

To seek observational evidence of star formation-powered submm emission
in high-redshift AGN, we 
and other groups have
carried out a sequence of surveys
of optically selected quasars at submm/mm wavelengths
(e.g. radio-quiet quasars at 
$z\sim$2: Priddey et al. 2003a and Omont et al. 2003; 
at $z>4$: Carilli et al. 2001, Omont et al. 2001, Isaak et al. 2002; 
and $z>5$: Priddey et al. 2003b; Petric et al. 2003; Robson et al. 2004). 
A handful of BALQs were incidentally observed during these studies,
exhibiting marginally higher detection rates compared with non-BAL quasars.
In addition, 
a number of the most spectacular high-redshift submm sources 
were also known to be BALQs: for example, the lensed 
$z$=2.55 (low ionization) BAL quasar H1413+117 known 
as the ``Cloverleaf'' (Hughes et al. 1997 for submm data), or
the luminous, lensed $z$=3.91 quasar APM08279+5255
(Lewis et al. 1998).
The implied link between BALs and large submm luminosity (hence possibly
star formation) could, if verified, provide
support for the evolutionary, ``unification by time'' model, 
but the sample sizes were
inadequate to confirm or refute this suggestion with statistical confidence.

Here we present a study of a sample of BAL quasars undertaken with
the SCUBA submm bolometer array on the JCMT.
As described below (Section 2.1), the sample has been selected to facilitate 
comparison with the benchmark sample of optically luminous,
$z\sim2$ quasars observed by Priddey et al. 2003a 
(hereafter P03). 
Explicitly, we wish to test the null hypothesis
that the submm properties of BALQs and non-BAL quasars are identical
(strictly, in this instance, that BALQs and non-BALs quasars have the same 
850$\mu$m flux distribution).
Ancillary goals of this work are: to determine accurately the typical 
submm properties of a well-selected, widely studied sample of BAL quasars, 
to establish the submm contribution to their broadband SEDs, and 
to search for dependence of submm flux on other observed
properties.

\section{SAMPLE SELECTION AND OBSERVATIONS}

\subsection{Sample selection \& strategy}
Our parent sample is derived from the catalogue of BALQs 
in the Large Bright Quasar Survey (LBQS: Hewett et al. 1995) 
identified by Weymann et al. (1991). 
This is a well suited sample for a number of reasons. The 
LBQS is a homogeneous, rigorously selected survey, and the BALQ
subset 
has formed the target of several detailed, multiwavelength studies 
(e.g. optical: Weymann et al. 1991, 
X-ray ({\it Chandra}): Gallagher et al. 2006a; 
mid-infrared ({\it Spitzer}): Gallagher et al. 2006b).
Also, the sample is well matched, in optical luminosity and redshift,
to the SCUBA Bright Quasar Survey (SBQS) $z$=2 sample (P03).
The LBQS was one of the sources for the $z$=2 SBQS;
note that all the BALQs in our sample meet the SBQS selection criteria.
(Indeed, some of them were incidentally observed during that project,
but will be removed from the P03 sample for the purposes of
defining a non-BAL quasar comparison sample.)
The correspondence in ($z$,$M_B$) between the two samples is illustrated
in Figure 1. 
Absolute $B$ band magnitudes ($M_B$) for the BALQs were extrapolated
(assuming a spectral index $-$0.5) from 
measurements of the monochromatic continuum flux at rest wavelength
$\approx$2000\AA\ measured by Weymann et al. (1991).
The definition of the original SCUBA (non-BAL) quasar surveys 
(optical luminosity criterion $M_B<-27.5$)
(McMahon et al. 1999, Isaak et al. 2002) employed
an Einstein--de Sitter cosmology with 
$\Omega_{\Lambda}$=0, $\Omega_{\rm M}$=1, $H_0$=50 km s$^{-1}$ Mpc$^{-1}$.
However, in Figure 1 and Table 1, we have transformed $M_B$ into 
a flat cosmology with 
$\Omega_{\Lambda}$=0.73 and %$\Omega_{\rm M}$=0.27, 
$H_0$=70 km s$^{-1}$ Mpc$^{-1}$ (Spergel et al. 2006), 
and the equivalent selection
criterion is shown as the dotted curve.
Over the redshift range $1.5<z<3$, the offset between the
cosmologies is small (indeed, their distance moduli are equal at
$z\approx1.5$).

The majority of the targets are {\it bona fide} radio-quiet quasars,
according to the definition of Stocke et al. (1992), who report
5GHz fluxes/upper limits for the majority of the sample.
1.4Ghz fluxes for the remainder were obtained from the FIRST (Faint Images of
the Radio Sky at Twenty centimetres: White et al., 1997) 
and NVSS (NRAO VLA Sky Survey: Condon et al. 1998)
catalogues. Only one object, LBQS B2211$-$1915 has a nearby (1.5 arcsec) 
radio source, with a 1.4GHz flux of 64mJy.
However, this quasar was not detected at 850$\mu$m so its radio-loudness
does not affect our analysis.
For the remainder, the radio upper limits predict a negligible ($\la$0.1mJy)
synchrotron component when extrapolated to the submm (e.g. using
a spectral index of $-$0.7, as in McMahon et al. 1999).

The final selection of observed sources was determined 
by largely uncontrollable factors such as telescope scheduling, etc.

\begin{figure}
\epsfig{figure=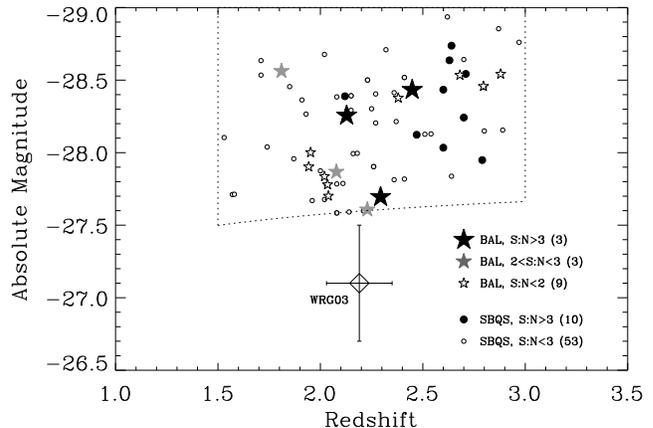,width=90mm}
\caption{Selection criteria for the SCUBA BALQ sample.
Star-shaped symbols indicate LBQS BALQs
observed with SCUBA:
$>$3$\sigma$ detections,
2$<\sigma<$3 tentative detections and $<$2$\sigma$ non-detections,
as illustrated by the key (number of members of each class given
in parentheses).
For comparison, plotted as circles are quasars from the SBQS survey
(Priddey et al. 2003a): $>$3$\sigma$ detections and non-detections (see key).
The selection area is plotted as a dotted line.
Finally, the large cross shows the mean $\pm$1$\sigma$ properties of the
sample of Willott, Rawlings \& Grimes (2003).}
\label{}
\end{figure}

In previous SCUBA surveys of high-redshift quasars, we adopted
a two-pronged strategy. For example in McMahon et al. (1999), 
a small $z>4$ sample was observed to a sensitive rms $\approx$1--1.5mJy
in order to establish accurately the 850 and 450$\mu$m 
properties of high-redshift quasars.
Isaak et al. (2002) and P03 (SBQS) pursued the complementary course
of observing larger samples ($\approx$40 and 60 respectively) 
to a shallower limit, 1$\sigma\approx$3mJy, with the primary aim of 
sifting out the bright detections as targets for future follow-up.
Since, however, the detection fraction at the limit
of the SBQS was only $\sim$25 percent, the ``typical'' submm flux of
the majority of $z$=2 quasars is clearly much fainter than this limit. 
The aims of the present BALQ programme 
(measure characteristic flux of BALQs, and precision photometry to constrain
SEDs) thus require high sensitivity, and so an rms of
1.5mJy at 850$\mu$m was adopted. 
In parallel with, and complementary to, the survey of P03, 
a set of more sensitive data (1$\sigma\approx$1mJy) 
was also obtained for a small number (5)
of (non-BAL) $z$=2 targets. Selection criteria, observing methods and
data reduction were equivalent to those employed in P03 ({\it q.v.}). 
(For reference, we summarise the 850$\mu$m fluxes of these objects here:
HS B0105+1619: 2.85$\pm$0.82mJy; HS B0218+3707: 0.95$\pm$1.17mJy; 
HS B0219+1452: 1.68$\pm$1.06mJy; HS B0248+3402: $-$0.49$\pm$1.08mJy;
HS B2245+2531: 2.66$\pm$0.91mJy; 
though we defer any more detailed discussion of these objects to a
separate paper (Priddey, Isaak \& McMahon, in prep.).)
These deeper data, in combination with the data from
P03 minus known BALQs, we subsequently refer to as the ``benchmark sample''.

\subsection{Observations \& data reduction}
Observations were carried out at JCMT through the periods
February--July 2003 and March--December 2004. 
Throughout the latter part of 2004, SCUBA experienced intermittent
technical problems involving a 3-day warm-up cycle in the cryogen
unit\footnote{See http://www.jach.hawaii.edu/JCMT/continuum/news.html
for details.}.
A number of the observations reported in this paper were taken during
this period, however the bulk of these data were obtained
during a spell of stable behaviour. 
We have carefully inspected the affected observations nevertheless,
to verify that the data appear to be of good quality.

The zenith opacity was monitored constantly via the JCMT
Water Vapour Monitor (WVM), and regularly through skydips.
These measurements were compared with each other and with 
the ``Tau Meter'' operated by the neighbouring Caltech Submm Observatory
(CSO).
Flux calibration was carried out, where possible, against
the planets Uranus and Mars, and when planets were not visible, by
using standard secondary flux calibrators. %(e.g. CRL618, ).

Data reduction was performed using the automated {\sc ORAC-DR}
(Jenness et al., 2002) 
pipeline and, independently, using the {\sc SURF} 
(Jenness \& Lightfoot, 1998) package.
Standard procedures were followed, involving: flatfielding, correction 
for atmospheric extinction, sky removal and clipping at the 3$\sigma$ level
to remove spikes. 

\subsection{Comparison of strategy with other SCUBA BALQ surveys}
An independent survey of the submm properties of BAL quasars
(with similar scientific motivation) was carried
out by Willott, Rawlings \& Grimes (2003; hereafter WRG03). 
Their strategy was complementary to the one pursued here, aiming for
a large (30 targets) sample observed to a relatively shallow 
(typically 2.5mJy rms) flux limit.
They found no statistically significant difference between their BALQ 
sample, and the sample of P03.
However, these two samples are not well matched, the SDSS-derived sample
of WRG03 lying $\approx$2 magnitudes fainter in the optical than the 
sample of P03 (e.g. Figure 1). 
Thus, a comparison between their BALQ sample and the non-BAL quasars 
sample of P03
involves making additional, poorly constrained assumptions about the
relation of submm with optical luminosity.
Moreover, their larger observational uncertainty in principle 
precludes detailed study 
of the detected sources (for example accurate measurement of SEDs).

\section{RESULTS AND ANALYSIS}
850$\mu$m and 450$\mu$m fluxes for the sample are shown in Table 1.
15 BALQs were observed in total, for all but one of which 
(LBQS B2211$-$1915) our completeness criteria were met, i.e. an 850$\mu$m
rms better than 1.5mJy or a detection better than 3$\sigma$.
Note that only one of the observed sample, LBQS B1231+1320, is a known 
low-ionization BAL quasar (LoBAL);
the others are all high-ionization BAL quasars (HiBALs). 
This is by happenstance rather than by design, the distinction between
the classes not having formed part of our selection criteria.
However, as LoBALQs make up $\sim$10 percent of optically selected 
BALQ samples, 1/15 is not unrepresentative.

\begin{figure}
\epsfig{figure=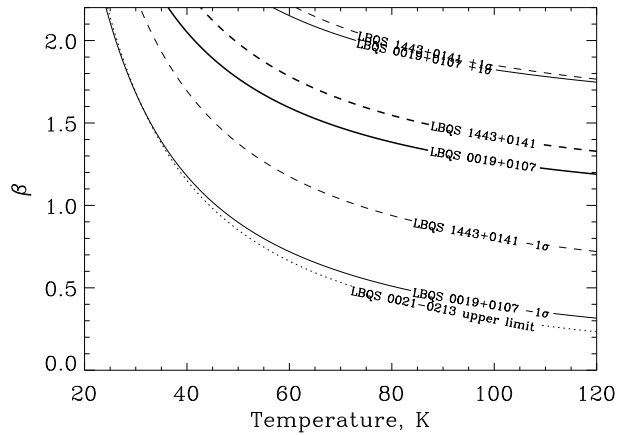,width=90mm}
\caption{Constraints on thermal dust SEDs for detected sources.
The curves trace loci of the observed 450/850$\mu$m ratios (thick lines) and
their $\pm$1$\sigma$ uncertainties (thin lines) or upper limit (thin 
dotted line for LBQS 0021$-$0213). For the 450$\mu$m-detected sources, 
either hot dust is required ($\sim$80K for a standard value for $\beta$
of 1.5), or a steep greybody spectral index $\beta\approx2$ (consistent with
$z>4$ quasar submm SEDs: Priddey \& McMahon 2001). Inevitably, 
the relative insensitivity at shorter submm wavelengths 
biases us toward being able to measure temperatures preferentially for
hotter sources.}
\label{}
\end{figure}

\subsection{Detected sources}
Three BALQs are detected with $>$3$\sigma$ significance at 850$\mu$m
(0019+0107, 0021$-$0213, 1443+0141). In addition, a further three
targets are tentatively detected at the 2--3$\sigma$ level of significance.
If we also include data from P03, then the total coadded 850$\mu$m flux
density for one of these marginal sources, LBQS B0025$-$0151, becomes 
3.8$\pm$1.2mJy--- confirming 
its detection at $>$3$\sigma$ significance. 
Note, however, that its flux lies below the formal
detection limit of the complete sample, 4.5mJy (3$\sigma$),
so it will not be taken into account in, for example, the detection rate
estimate (below).

Two sources are also detected at 450$\mu$m, with $>$3$\sigma$ significance
(0019+0107 and 1443+0141). 
Although the 
450$\mu$m rms errors are large 
(combined with which
one must also take into account a calibration error of $\sim$20 percent,
when converting from a signal in volts to a flux density in mJy),
the 850/450$\mu$m flux ratios may be used to make crude estimates of 
the temperature of the emitting dust.
Figure 2 shows the constraints placed on an isothermal dust SED
parametrized by temperature and greybody index $\beta$. 
The two sources detected at 450$\mu$m require hot dust ($\ga$80K)
if a typical value of 1.5 is assumed for $\beta$. However,
for $\beta$=2, the temperatures descend to $\approx$40K---
consistent with submm SEDs determined for $z>4$ quasars by Priddey
\& McMahon (2001). 
In contrast, a firm upper limit can be placed on the 450/850$\mu$m
ratio of LBQS B0021$-$0213, placing a 2$\sigma$ upper limit on
the dust temperature of $\la$40K.

Further multiwavelength data are needed to break the $T-\beta$
degeneracy, and thereby determine the origin of the submm emission.
A low dust temperature would be suggestive of emission not from
the central regions of the AGN, but from
the kpc scales of the host galaxy, possibly from regions of
massive star formation. Taking $T$=40K, $\beta$=2.0,
the integrated thermal luminosity, in the far-infrared, of 
LBQS B0019+0107 is $L_{\rm FIR}$=1.6$\pm$0.5$\times$10$^{13}$L$_{\odot}$,
translating to an instantaneous rate of massive stars of
$\sim$2000M$_{\odot}$ yr$^{-1}$ 
(e.g. following the same method and assumptions
as in Isaak et al. (2002)).
On the other hand, a hot temperature would be expected from
dust in close proximity to the AGN, lying either in a ``unified scheme'' 
torus, or associated with the BAL wind.
Taking $T$=80K, $\beta$=1.5, the far-infared luminosity of 
0019+0107
is $L_{\rm FIR}$=1.3$\pm$0.4$\times$10$^{14}$L$_{\odot}$--- comparable with
the bolometric luminosity $L_{\rm bol}\approx$2$\times$10$^{14}$L$_{\odot}$
inferred from the optical (e.g. using the $B$ band bolometric correction
factor of Elvis et al. 1994).
Observations with the MIPS and IRAC instruments on {\it Spitzer}, which
could shed light on the question of the location of the dust
and the origin and energetics of the dust-heating light,
have been obtained
for all of the current sample, and these will be discussed in
a forthcoming paper (Gallagher et al. 2006b).

\begin{table*}
\caption{Submm fluxes of the LBQS BALQ sample. Statistically
significant detections ($>3\sigma$) are shown in bold.}
\begin{tabular}{lllcrr}
Source name &  $z$ & $B_{\rm J}$ & $M_B$ & $S_{850}$ & $S_{450}$ \\
            &  & &     &  (mJy)     & (mJy) \\
\hline
LBQS B0019+0107 & 2.130& 18.09 & $-$28.3 & {\bf 8.2$\pm$2.3} & {\bf 50$\pm$16} \\
LBQS B0021$-$0213 & 2.293& 18.68 & $-$27.7 & {\bf 5.3$\pm$1.1} &3.8$\pm$6.8 \\
LBQS B0025$-$0151 & 2.076& 18.06 & $-$27.9 & 3.5$\pm$1.4 & $-$2$\pm$16 \\
\multicolumn{4}{c}{+ additional data from Priddey et al. (2003)}
& {\bf 3.8$\pm$1.2} & \\
LBQS B0029+0017 & 2.253& 18.64 & $-$27.6 & 5.2$\pm$2.0 & 19$\pm$11\\
LBQS B1029$-$0125 & 2.029& 18.43 & $-$27.7 & 1.0$\pm$1.2 & 4.4$\pm$6.4 \\
LBQS B1208+1535 & 1.961& 17.93 & $-$27.9 & 2.2$\pm$1.4 & 16$\pm$9 \\
LBQS B1231+1320 & 2.380& 18.84 & $-$28.4 & $-$0.1$\pm$1.4 & 1.3$\pm$7.5 \\
LBQS B1235+0857 & 2.898& 18.17 & $-$28.5 & 1.2$\pm$1.1 & $-$7.5$\pm$6.9 \\
LBQS B1235+1453 & 2.699& 18.56 & $-$28.5 & 1.9$\pm$1.4 & $-$10$\pm$20 \\
LBQS B1239+0955 & 2.013& 18.38 & $-$27.8 & $-$0.7$\pm$1.0 & $-$5.7$\pm$5.4 \\
LBQS B1243+0121 & 2.796& 18.50 & $-$28.5 & 1.6$\pm$1.1 & $-$4.6$\pm$7.2 \\
LBQS B1443+0141 & 2.451& 18.20 & $-$28.4 & {\bf 5.2$\pm$1.2} & 
{\bf 33.8$\pm$8.5}\\
LBQS B2154-2005 & 2.035& 18.21 & $-$27.8 & 2.2$\pm$1.5 & 10$\pm$11 \\
LBQS B2201$-$1834 & 1.814& 17.81 & $-$28.6 & 3.0$\pm$1.5& $-$25$\pm$13 \\
LBQS B2211$-$1915 & 1.952& 18.02 & $-$28.0 & 0.8$\pm$1.65 & $-$19$\pm$19 \\
\hline
\end{tabular}
\flushleft
{\bf Notes:}
$z$, $B_{\rm J}$ from Hewett et al. (1995)
\end{table*}

\subsection{Comparison with non-BAL quasars}
Figure 1 illustrates the correspondence in selection criteria 
between the $z\sim2$ sample from P03, and the final BALQ sample
observed here. 

Considering the SCUBA-observed BALQ sample,
out of 14 targets observed down to the completeness limit ($\sigma$=1.5mJy),
3 are detected at $>$3$\sigma$ significance. This gives us
a detection fraction of 0.21$\pm$0.12.
If we set the detection threshold at $>$2$\sigma$, the fraction
becomes 0.43$\pm$0.17.
Using a fit to the flux density distribution of the benchmark sample
(see below for details), a lognormal function with 
median flux 2.00mJy and standard deviation 0.38dex, 
we predict a 3$\sigma$ (2$\sigma$) 
detection fraction, for a survey 
attaining 1.5mJy rms, of 0.21 (0.42) (Table 4)---
closely consistent with the observed detection fraction of BALQs
(notwithstanding the large uncertainty in the latter).
Thus from a simple detection rate analysis, it appears that our data 
confirm the 
finding of WRG03, that
BALQs as a class are not significantly brighter in the submm than the general
$z$=2 radio-quiet quasar population. 
In the next two sections we investigate more detailed methods of comparison.

\subsection{Survival analysis}
The statistical methods of {\it survival analysis} can be used to take into
account information from upper limits--- or ``left censored data''.
To examine the null hypothesis $\rm{H_0}$ that BALQs and non-BAL quasars have
the same submm flux density distribution, we have carried out the tests
described by Feigelson \& Nelson (1985) 
(using the software package {\sc ASURV}, La Valley et al. 1992), 
enabling intercomparison of
two censored datasets.
The tests return the significance level at which $\rm{H_0}$ may be rejected.
Table 2 shows the results of comparing the BALQ samples against the
benchmark sample. 
Mean fluxes obtained from the censored data using the Kaplan--Meier
estimator, along with significance levels corresponding to the comparison 
with the benchmark sample, are also given.
Two conventions for discriminating between detections and non-detections
are employed: 
A. signal-to-noise$\ge$3$\longrightarrow$detection; 
B. signal-to-noise$\ge$2$\longrightarrow$detection.
(Note that the latter convention is used by WRG03.)
We employ a conservative estimate of the value of the upper limit,
corresponding to $S_{\rm ul}$=max($S$,0)+2$\sigma$, 
rather than the alternative $S_{\rm ul}$=2$\sigma$. 

\begin{figure}
\epsfig{figure=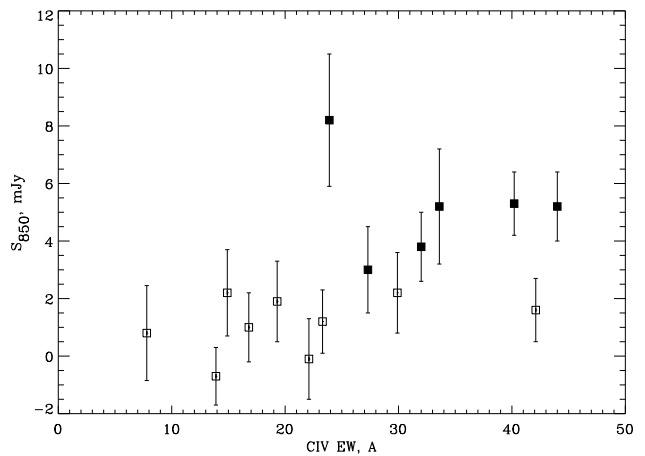,width=90mm}
\caption{850$\mu$m flux density plotted against equivalent width of the
C {\sc IV} broad absorption line. Submm detections ($>2\sigma$) are plotted as
filled squares; non-detections as open squares (as points, at the
value of their measured fluxes $\pm$1$\sigma$). 
It is apparent that large EWs ($\ga$25\AA) dominate the detected sources:
the K--S probability that the detections and non-detections have the
same EW distribution is $\approx$0.01.}
\end{figure}

\begin{table*}
\caption{Survival analysis of the three samples.
The table is split according to whether a signal-to-noise ratio greater than
3 or 2 is taken to imply a detection (A and B respectively).
$p$ is the probability of the null hypothesis that
the samples have the same 850$\mu$m flux distribution.
These tests are carried out between the BALQ samples and the benchmark
$z$=2 sample (see text).}
\begin{tabular}{l|llll}
Sample: & benchmark & WRG03 & LBQS & WRG03+LBQS \\
       & $z$=2 non-BAL & faint BALQs & bright BALQs & all BALQs \\
\hline
\multicolumn{5}{c}
{\underline{A. Censored $\longleftrightarrow$ S:N $<$3}}\\
Mean & 3.72$\pm$0.48 & 5.06$\pm$0.36 & 3.72$\pm$0.48 & 3.42$\pm$0.44 \\
$p$ logrank & | & 0.99 & 0.70 & 0.83 \\
$p$ Gehan & | & 0.93 & 0.48 & 0.77 \\
$p$ Peto--Peto & | & 0.97 & 0.62 & 0.83 \\
$p$ Peto--Prentice & | & 0.97 & 0.67 & 0.84 \\
\\
\multicolumn{5}{c}
{\underline{B. Censored $\longleftrightarrow$ S:N $<$2}}\\
Mean & 4.52$\pm$0.47 & 5.11$\pm$0.35 & 4.52$\pm$0.47 & 3.15$\pm$0.62 \\
$p$ logrank & | & 0.96 & 0.21 & 0.30 \\
$p$ Gehan & | & 0.47 & 0.11 & 0.16 \\
$p$ Peto--Peto & | & 0.78 & 0.14 & 0.26 \\
$p$ Peto--Prentice & | & 0.74 & 0.16 & 0.24 \\
\hline
\end{tabular}
\end{table*}

Employing several estimators to perform the comparison (see Feigelson
\& Nelson 1985 for details), Table 2 shows that
the null hypothesis, that BALQs and non-BAL quasars have identical submm flux
distributions, cannot be rejected with any strong degree of confidence
(i.e. at a significance level $p\la0.05$). 
The largest disparity, according to this method, occurs between the 
SCUBA-observed LBQS BALQ sample 
and the benchmark sample, but since this is at best only 
$\approx$10 percent probable, the result is only marginal, and
could be an artefact of the deeper limit reached in the LBQS sample. 

We can also use bivariate survival analysis (Isobe, Feigelson \& Nelson, 1986) 
to test for correlations between
submm flux and other 
observables\footnote{The relevant quantities are conveniently tabulated
for our sample by Gallagher et al. 2006a.}. 
Censoring at the 2$\sigma$ level,
there appears to be a marginally significant 
correlation between submm flux and the equivalent width (EW) of the 
characteristic C {\sc IV} broad absorption line 
($p\equiv$probability that no correlation is present$\approx$0.04: Table 3).
Figure 3 illustrates that six of the eight largest EWs correspond
to submm detections; whilst below $\approx$25\AA\ there are no detections
at all.
Indeed, the Kolmogorov--Smirnov (K--S) two-sample test returns
a probability of 0.01 that the detections and non-detections have the
same C {\sc IV} EW distribution--- a statistically significant result.

On the other hand, the Balnicity Index (BI: Weymann et al. 1991), 
a more conservative
measure of the C {\sc IV} EW (counting only high-velocity ($>$3000km s$^{-1}$)
absorption above 2000 km s$^{-1}$ in width),
returns a somewhat lower significance for the correlation test
($p\approx0.09$: Table 3).
Although the BI is well correlated with CIV EW for the LBQS BAL sample
as a whole,
the discrepancy could be explained by a large scatter in this correlation
($\sim$10 percent), which increases (to $\sim$25 percent) at low
equivalent widths (below $\sim$25\AA)--- smearing out any effect such as
that seen in Figure 3.
Similarly, submm flux against the optical--X-ray spectral index
$\alpha_{\rm ox}$, which is itself correlated with the C {\it IV} EW
for low-redshift 
($<$0.5) radio-quiet quasars (including a handful of
BALQs) (Brandt, Laor \& Wills, 2000), returns no hint of a correlation.
For this reason, combined with its reliance upon 
a 2$\sigma$ detection threshold, we consider the 
$S_{850}$--C {\sc IV} result extremely suggestive, but not,
without further observation,
conclusive.

\begin{table}
\caption{Tests for correlations between 850$\mu$m flux and 
other observables of the sample.
$\alpha_{\rm uv}$ is the ultraviolet continuum slope, and
$\Delta\alpha_{\rm ox}$ is a proxy for the amount of X-ray absorption
(see Gallagher et al. 2006a).
$p_{corr}$ is the probability that no correlation is present;
$p_{KS}$ is the Kolmogorov--Smirnov probability that the detections
and non-detections have the same distribution of the variable.
(N.B. the Spearman's Rho estimator has {\it not} been used because it is deemed
to be unreliable for small datasets)%
}
\begin{tabular}{lllll}
Variable & C IV EW  & BI & $\alpha_{\rm uv}$ & $\Delta\alpha_{\rm ox}$\\
\hline
&\multicolumn{4}{c}{$p_{\rm corr}$}\\
Proportional Hazard & 0.037 & 0.082 & 0.35 & 0.22 \\
Kendall's Tau & 0.041 & 0.095 & 0.50 & 0.11 \\
&\multicolumn{4}{c}{$p_{\rm KS}$}\\
K--S two-sample & 0.011 & 0.36 & 0.98 & 0.53 \\
\hline
\end{tabular}
\end{table}

\subsection{Flux density distribution--- Bayesian method}
It is noteworthy that the vast majority of the 850$\mu$m fluxes
are positive (with a mean significantly different from zero), 
despite careful sky removal.
In contrast, 450$\mu$m fluxes are both negative and positive, as
one would 
expect for an average signal consistent with zero.
This effect was scrutinized in detail in our previous
submm photometric surveys, e.g. Isaak et al. (2002), where a
careful comparison between on- and off-source bolometers
revealed a net positive, on-source flux.
A reasonable interpretation is that, at 850$\mu$m, 
we are measuring a real, non-zero flux distribution---
a distribution with a positive characteristic flux density
(e.g. the ``fiducial'' flux discussed by McMahon et al. 1999)
and a width reflecting an intrinsic spread---
which has been further smeared out by observational noise.
This poses a problem, {\it inter alia}, 
as to what to do with fluxes lying at the 2--3$\sigma$ level:
whether or not to consider them as ``detections'', and how much
confidence to attach to the measured signals.
At low signal-to-noise, there is the risk of artificially boosting
fluxes if the effects of Malmquist bias are not properly taken into 
account; on the other hand, setting too stringent a detection threshold
discards potentially useful information about the underlying flux
distribution.

To quantify these effects, and to estimate the range of true
submm flux distributions that are consistent with our observations,
we invert the problem via a Bayesian technique--- starting from a 
range of models and iteratively updating the probability distributions
over their parameters by comparison with each observed source.
Details of this technique are described in Appendix A;
in companion papers (e.g. Priddey, Isaak \& McMahon, in prep.) we
explore its application to a wider range of submm datasets than
required for the current BALQ analysis.

Taking a lognormal distribution as an illustrative example,
contours of the final probability distribution over its
parameters--- median ($S_0$) and standard deviation ($\Sigma_{10}$:
in dex units,
denoted by upper case $\Sigma$ to distinguish it from the rms 
submm flux density $\sigma$)---
are shown in Fig 4. 
Also plotted is the marginal distribution for $S_0$ 
(calculated by integrating over the final joint probability
distribution $P(S_0; \Sigma_{10})$ with respect to $\Sigma_{10}$).
$S_0$ provides a formal measure of the 
``fiducial flux'' of McMahon et al. (1999), the characteristic
flux density of the underlying distribution.
For the LBQS BALQ sample, the most probable
value for $S_0$=2.02mJy, and 
for $\Sigma$ it is 0.30dex\footnote{Note, the width of this function
is a measure of the intrinsic spread of the underlying distribution, 
it is {\it not} a measurement uncertainty.}.
The best fit distribution is plotted as differential and cumulative
histograms, together with the data. In these figures, the plotted
model distribution does {\it not} directly represent the {\it underlying} 
distribution, but has been smoothed so as to represent the 
{\it observed} distribution that would be obtained 
if all observations had the same rms flux (equal in this case
to the mean rms of the sample).
Note, however, that the parameter estimation 
method is more subtle than this, taking full account of individual
rms values (i.e. it is not as simple as deconvolving an observed histogram
with a constant rms).

As noted above (Section 3.2),  
we can also apply the same analysis to the benchmark $z$=2 quasars 
and to the WRG03 samples (Table 4).
Although the datasets employ complementary strategies
(e.g. broad-but-shallow versus deep-but-narrow), 
the Bayesian method takes account of the differing rms values,
making it possible to combine the two BALQ samples within a single analysis. 
Table 4 shows that the samples are, on the whole, consistent with 
each other, supporting the conclusions of the other statistical 
techniques we have used in this paper. 

We have also applied the procedure to the 450$\mu$m fluxes of the current
LBQS BALQ sample (the other samples do not have adequate 450$\mu$m data).
In this case, the weighted mean is not significantly different from zero
(though it is positive, as expected given the two statistically significant
450$\mu$m detections),
and although fits are obtained, their quality is poorer than for the
850$\mu$m samples 
(N.B. only the Gaussian distribution allows for a {\it negative} 
characteristic flux, to act as a ``sanity check'' on the method
in case of overall null detections: 
the other distributions assume that in the absence of noise, real astronomical
objects have positive fluxes!).

\begin{figure}
\epsfig{figure=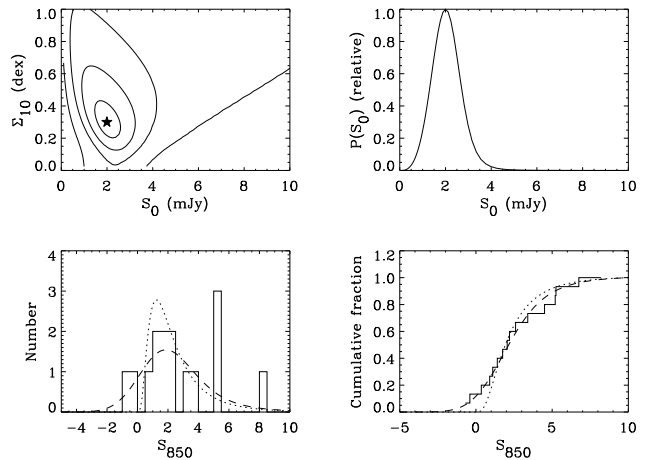,width=90mm}
\caption{Lognormal fit to the 850$\mu$m flux distribution of the LBQS
BALQ sample. {\bf Top left}, log likelihood contours over the lognormal
parameters. The model with the highest likelihood is plotted with a
star symbol, and contours correspond to (1,2,3,4)$\sigma$.
 {\bf Top right}, marginal distribution over median flux
(probability relative to peak).
Models where $S_0\approx2$mJy are strongly favoured.
{\bf Bottom left}, histogram of observed fluxes, with the underlying
best-fit distribution plotted as a dotted curve. The dashed curve 
simulates the distribution that would actually 
be observed, given the best fit, assuming a constant rms is obtained
(equal to the observed sample average in this case) 
{\bf Bottom right}, as bottom left, but plotted as a cumulative 
histogram.}
\end{figure}

\begin{table*}
\caption{Properties of the 850$\mu$m flux distributions of $z\sim2$
BALQ and non-BAL quasar samples (and 450$\mu$m distribution of the LBQS BALQ
sample: the others do not have adequate 450$\mu$m data). 
The first four rows are non-parametric
statistical measures (average rms, then mean, median and 
variance-weighted flux density), the remainder describe the Bayesian estimates
of the underlying distributions: lognormal, Schechter and Gaussian
distributions (see Appendix A, Equations A1--A3 for functional forms). 
The BALQ sample from this paper is 
designated ``LBQS''. $f_{\rm det}(>4.5)$ and ($>3.0$) are the predicted
3$\sigma$ (2$\sigma$) detection fractions assuming a 
(consistent) survey limit 1.5mJy.}
\begin{tabular}{r|lllll}
Sample: & benchmark & WRG03 & LBQS & WRG03 + LBQS & LBQS \\
       & $z$=2 non-BAL & faint BALQs & bright BALQs & all BALQs & (450$\mu$m)\\
\hline
$<\sigma>$ & 2.67 & 2.53 & 1.40 & 2.15 & 10.9 \\
$<S>$ & 2.89 & 2.56 & 2.70 & 2.62 & 4.3 \\
$S_{med}$ & 2.33 & 3.45 & 2.20 & 2.20 & 1.3 \\
$<S>_{w}$ & 2.53$\pm$0.29 & 2.55$\pm$0.45 & 2.28$\pm$0.34 & 2.41$\pm$0.27 & 
3.2$\pm$2.2\\
\\
lognormal $S_0$ & 2.00$\pm$0.50 & 2.02$\pm$0.65 & 2.02$_{-0.60}^{+0.50}$ 
& 2.05$\pm$0.45 & 1.4$_{-1.1}^{+2.6}$\\
$\Sigma_{10}$ & 0.38$_{-0.07}^{+0.10}$ & 0.33$_{-0.11}^{+0.15}$ 
& 0.30$_{-0.10}^{+0.13}$ & 0.31$^{+0.10}_{-0.07}$ & 0.85$_{-0.3}^{+1.0}$ \\
$f_{\rm det}(>4.5)$ & 0.21 & 0.20 & 0.18 & 0.19 & | \\
$f_{\rm det}(>3.0)$ & 0.42 & 0.41 & 0.40 & 0.41 & | \\
\\
Schechter $S_0$ & 2.9$\pm$1.3 & 1.98$^{+1.8}_{-1.0}$ & 1.26$_{-0.6}^{+1.4}$ 
& 1.60$_{-0.60}^{+1.0}$ & 15$_{-7}^{+9}$\\
$\alpha$ & 0.0$_{-0.3}^{+0.5}$ & 0.33$_{-1.0}^{+1.5}$ & 1.00$^{+2.0}_{-1.0}$ 
& 0.60$_{-0.8}^{+1.0}$ & $-$0.5$_{-0.1}^{+0.3}$\\
%$f_{\rm det}(>4.5)$ & 0.20 & 0.15 & 0.13 & 0.13 \\
$f_{\rm det}(>4.5)$ & 0.25 & 0.22 & 0.20 & 0.21 & | \\
$f_{\rm det}(>3.0)$ & 0.45 & 0.43 & 0.42 & 0.43 & |\\
\\
Gaussian $S_0$ & 2.74$\pm0.5$ & 2.53$\pm$0.6 & 2.49$^{+0.6}_{-0.5}$ 
& 2.53$\pm$0.45 & 2.5$\pm$2.5 \\
$\Sigma$ & 2.53$^\pm$0.45 & 2.34$\pm$0.65 & 1.61$^{+0.5}_{-0.4}$ 
& 1.94$\pm$0.45 & 3.3$^{+1.7}_{-2.0}$ \\
%$f_{\rm det}(>4.5)$ & 0.24 & 0.20 & 0.11 & 0.15 & | \\
$f_{\rm det}(>4.5)$ & 0.32 & 0.28 & 0.20 & 0.24 & | \\
$f_{\rm det}(>3.0)$ & 0.55 & 0.52 & 0.46 & 0.49 & | \\
\hline
\end{tabular}
\end{table*}

\section{DISCUSSION}
Both the geometry and the lifetime of BAL outflows are poorly constrained,
to the extent that the observed fraction of BALQs amongst the general
quasar population ($\sim$15 percent), could be accounted for either
as an effect of orientation, of evolution, or of some combination of
both.
In the geometric interpretation, an anisotropic outflow, such as
an accretion disk wind (e.g. Murray et al. 1995), 
is believed to be a feature of all AGN 
which is, however, only visible along a narrow range of sightlines.
In this model, whether the observer sees a Type 1 (broad-line, unabsorbed)
quasar, Type 2 (narrow-line, absorbed) quasar or BALQ (broad-line absorbed), 
depends on viewing angle.
The contrasting evolutionary model holds that BALs occur when 
the AGN has been recently (re)fuelled, cocooned within 
a dense cloud of gas and dust, possibly coinciding with 
a major starburst.
The BAL outflows may, furthermore, play a role in terminating
further accretion and star formation, marking the transition between
obscured AGN and classical, optical- and soft X-ray-luminous quasar.

Evidence for and against the evolutionary as opposed to the orientation
model is, currently, somewhat mixed. 
Similar optical emission-line and continuum properties suggest that
BALQs and non-BAL quasars belong to the same class 
(Weymann et al. 1991; Reichard
et al. 2003), and results from spectropolarimetry 
favour an equatorial viewing plane for BALQs (Ogle et al. 1999).
In contrast, radio spectral studies of  both radio-loud
(Becker et al. 2000) and radio-quiet (Barvainis \& Lonsdale 1997) BALQs
show that BALQs exhibit a mixture of radio spectral indices, 
both flat (core-dominated, pole-on) as well as steep (lobe-dominated, edge-on),
indicating a range of observed lines of sight. 

On which side of the fence do our results sit?
At face value, the observed similarity between BALQs and non-BAL quasars 
in the submm
would suggest that we are indeed observing intrinsically identical populations.
This conclusion was favoured by WRG03, and our deeper limits and more rigorous
selection would seem to lend it greater force.
However, we have demonstrated (for example by our Bayesian derivation of
the flux density distribution) that $z$=2 quasars, BAL and non-BAL alike, 
are typically very faint in the submm ($\sim$2--3mJy). 
The differences between BALQs and non-BAL quasars, if any, presumably lie
further down the luminosity function than current (single aperture) 
submm surveys, with relatively bright confusion limits, can reach.

On the other hand, our tentative finding of a submm dependence on C {\sc IV} EW
(a large EW seems to be a necessary, but not sufficient, condition
for submm detectability, since two high EW BALQs are nevertheless not
detected: Figure 3) suggests that yet more rigorous selection criteria
would yield substantially higher submm detection rates compared with non-BAL
quasars. 

It is of course possible that this effect results from selection biases,
for example due to reddening.
The most reddened quasars must be intrinsically more luminous to meet the 
$B_J$ selection criterion of the LBQS.
If the C {\sc IV} EW were, for whatever reason, correlated with reddening, 
this bias could induce the apparent correlation with submm 
flux\footnote{Such an effect would, however, be difficult to test, 
because it is difficult to determine the underlying continuum for an
individual BAL quasar (e.g. Trump et al. 2006).}.
Nevertheless, assuming that our sample is homogeneous in terms of
intrinsic luminosity, and that our submm--C {\sc IV} EW relation is valid,
what are its implications? 

\medskip
\noindent
It is well known that BALQs are weaker in X-rays, relative to the 
optical, than non-BAL quasars---
an effect believed to be due to absorption rather than an unusual 
intrinsic spectrum (e.g. Green et al. 1995; Gallagher et al. 2006a).
Indeed, Brandt, Laor \& Wills (2000) discovered,
for a sample of low-redshift ($z<0.5$) quasars, a correlation 
between their C {\sc IV} EW and soft X-ray weakness as measured by the
optical--X-ray continuum (2500\AA--2 keV) spectral index ($\alpha_{\rm ox}$).
This connection between UV and X-ray absorption does not hold so
straightforwardly for $z\sim2$ LBQS quasars (Gallagher et al. 2006: Fig. 6a).
Nevertheless, if our $S_{850}$--C {\sc IV} EW connection is real, it
could loosely be interpreted to advocate
a link between submm activity, UV absorption and
X-ray absorption. The link need not be direct, however, since 
the components responsible for each effect are likely to be distinct
(this could account for the findings of Gallagher et al. 2006a).
For example, X-ray absorption is an important feature of
radiation-driven wind models, in which a ``shielding gas'' component
is proposed, without which the outflow would become overionized
by soft X-rays, and 
UV resonance-line-driven radiation pressure would become ineffective.
The submm-emitting region, in contrast, 
is likely to be much more extended, lying within
for example a dust torus or the host galaxy.

A possible analogy could be drawn with Narrow Line Seyfert 1 (NLS1) galaxies.
AGN of this class 
are believed to result from SMBHs with an accretion rate that
is a high fraction of the maximal Eddington-limited rate,
and high velocity outflows have been observed in some
(e.g. Pounds et al. 2003). 
It is possible that every young AGN--- either freshly formed or recently
refuelled following a merger--- passes through such
a high-Eddington state, launching an outflow, 
when gas supply to the nucleus is plentiful. Accompanying star formation
could also result from, or play an active role in
(e.g. Thompson, Quataert \& Murray, 2005), 
the processes that drive gas into the nucleus.
As the AGN luminosity increases as the SMBH grows, 
radiation pressure can generate BAL outflows of sufficient magnitude to 
terminate both star formation and further SMBH accretion.

Page et al. (e.g. 2004) found a similar submm dependence on X-ray absorption
for a sample of $z\sim$2 quasars.
In contrast with the canonical geometric unified schemes for AGN
(e.g. Antonucci 1993),
this is interpreted as evidence of an evolutionary sequence, rather
like the schemes outlined above, between an initial, obscured, AGN--starburst
phase, and a final unobscured phase, in which star formation has ceased. 
A BAL wind could act as the transitionary mechanism between these two phases.

An emerging point of view is that it is in fact the {\it low-ionization} BALQs
that form a class distinct from other quasars (and from the more common
HiBALs), perhaps a manifestation of youthful AGN, with a high covering
factor of absorbing gas and perhaps dust 
(Boroson \& Meyers, 1992; Voit, Weymann \& Korista, 1993), and, indeed,
the optical continua of LoBALs seem redder than those of other
classes (Sprayberry \& Foltz 1992; Reichard et al. 2003).
One might expect a high detection fraction
in the submm and infrared if that were the case, particularly if this
youthful phase were accompanied by a starburst. 
The distinction between low- and high-ionization BALQs did not form
part of our selection criteria. The single (known) LoBAL in our sample, 
LBQS B1231+1320, was not detected ($S_{850}$=$-$0.1$\pm$1.4mJy). 
However, it would still be of interest to obtain submm/mm observations
of a sample of LoBALs, or the third, rarer subclass 
known as FeLoBALs (e.g. Becker et al. 1997).

\medskip
\noindent
In contrast to a starburst power source for the dust,
the submm component could derive from optical/UV AGN 
emission reprocessed by nuclear dust. In this case, (i) larger C IV EW
may merely be indicative of an intrinsically more luminous central engine.
In support of this idea, Brandt, Laor \& Wills (2000) found tentative 
evidence for a correlation between absolute visual magnitude ($M_V$) and 
C IV EW, though 
SCUBA and MAMBO samples (Omont et al. 2001; Isaak et al. 2002; 
Priddey et al. 2003; Omont et al. 2003)
do {\it not} show,
over the limited range of optical luminosities investigated in the
present BALQ sample,
such a strong correlation between optical and submm.
Alternatively, (ii) if all AGN exhibited a {\it distribution} of wind
covering fractions,
then those that fill a larger solid angle of the sky 
would be more likely both to 
show high far-infrared luminosity
(dust in the wind reprocesses a larger fraction of the AGN's UV/optical
output) and to exhibit BAL outflows 
(since a greater proportion of lines of sight intersect the wind).

Comparison of the submm fluxes and limits with the full SEDs 
(particularly X-ray and mid-infrared) will help constrain such 
speculation by casting light on the
starburst versus AGN origin of the submm emission.
But in either case, the implication of a dependence of submm flux
on the properties of UV absorption lines 
provides support for the idea that
the BAL phenomenon is not a {\it simple} geometric effect arising
from a wind with a fixed covering fraction:
since the X-ray- and UV-absorbing components are optically thin
to submm radiation, the submm properties would not, 
{\it all other factors being equal}, be expected to depend
on the observer's viewing angle relative to the wind. 
It is, of course, quite possible that a non-trivial combination of variables
(a range of lifetimes, covering factors and intrinsic luminosities)
must be invoked to unify the BALQ class within the diverse ensemble of 
AGN types.

\section{SUMMARY \& CONCLUSIONS}
We have observed a sample of $z$=2, optically luminous 
Broad Absorption Line quasars at submm wavelengths, using the 
SCUBA bolometer camera on the JCMT.
Our goals were twofold:

(i) to test the null hypothesis $\rm{H_0}$: {\it BALQs and non-BAL
quasars have identical submm properties}

(ii) to determine the 850$\mu$m flux distribution, the 
submm (850 + 450$\mu$m) contribution to the SEDs, and the relation
between submm and other properties, of a well-selected,
well-studied sample of BALQs.

The sample was selected so as to match, in redshift and optical
luminosity, the ``benchmark'' 
submm sample of optically luminous,
radio-quiet quasars observed by Priddey et al. (2003a) 
(supplemented with a small amount of additional data presented here).
A direct comparison between BALQ and non-BAL quasars is thus possible.
We have carried out this comparison in a number of different ways:
1. Simple non-parametric statistics such as means and medians;
2. Survival analysis;
3. A Bayesian method to disentangle the underlying flux
distribution (once a suitable functional form is assumed) 
from the observed distribution.
All these methods indicate that the submm properties of the BAL and 
non-BAL samples appear to be similar: $\rm{H_0}$ cannot be rejected
with confidence.
This conclusion is true for the luminosity-matched LBQS sample
presented in this paper, as well as for the combined LBQS sample
and that of WRG03 which, overall, spans a much larger range in optical
luminosity. 
Note, however, that the LBQS sample generally
shows larger deviations from the benchmark 
than the sample of WRG03 (though at best this effect is present
at the $\sim$10 percent level). 
There is, moreover, some indication that optically fainter, non-BAL
quasars at $z\sim$2 are also fainter in the submm
(Priddey, Isaak \& McMahon, in prep.).
A comparison between the WRG03 sample and the optically faint $z$=2 sample 
would tentatively indicate that BALQs were systematically {\it brighter} 
in the submm than non-BAL quasars, but detailed characterisation and
intercomparison of these two samples is limited by their large number of 
high rms, low-significance measurements.

The ``typical'' 850$\mu$m flux density of an optically luminous, $z$=2 
BALQ is $\approx$2--3mJy (estimated by non-parametric methods and via 
the Bayesian fit to the flux distribution). 
Thus only with very deep submm/mm observations 
will it be possible to detect the majority of the BALQ population
(when the difficulty will be encountering the confusion limit
of current-generation, single-dish submm telescopes),
and to carry out fully indicative tests of their submm properties
relative to the general quasar population.

Some 450$\mu$m detections/upper limits enable crude estimates of the 
dust temperature.
For a fuller discussion of the broadband SEDs (X-ray, optical, 
mid-infrared) of the LBQS BALQ sample, see Gallagher et al. (2006a)
and Gallagher et al. (2006b). In principle, observations with
{\it Spitzer} could discriminate 
spectrally
between nuclear (e.g. AGN torus)
and extended (e.g. star formation in host) dust components
(for example by fitting template SEDs appropriate for each component).

Finally, we find tentative evidence for a dependence of submm flux on
the equivalent width of the C IV broad absorption line.
Although difficult to establish definitively with the present data,
this finding would be easy to test by carrying out mm/submm
observations of samples selected according to the EW criterion
(specifically it would predict a high detection rate above 
EWs of $\approx$25\AA).
If this finding is correct--- and whether the dust is heated by
starburst or AGN--- it suggests that the BAL phenomenon is
not a simple geometric effect arising from an orientation-dependence
on a wind with a fixed covering fraction, 
but that other variables, such as
evolutionary phase, absorber covering fraction
or intrinsic luminosity,
must be invoked in order to unify BALQs with other classes of AGN.

\section*{Acknowledgments}
RSP gratefully acknowledges support from the University of Hertfordshire.
Support for SCG was provided by NASA through the Spitzer Fellowship 
Program, under award 1256317.
The authors thank the referee, Alexandre Beelen, for comments.
The JCMT is operated by the Joint Astronomy Centre in Hilo,
Hawai`i, on behalf of the parent organisations: the Particle Physics and
Astronomy Research Council in the United Kingdom, the National Research
Council of Canada and the Netherlands Organisation for Scientific Research.

%%%''These were the first made of his creatures: their hearts were of fire, 
%%%but they were cloaked in darkness, and terror went before them; 
%%%they had whips of flame. BALrogs they were named...''

%%%%%APPENDIX ADDED TO ADDRESS REFEREE'S AND EDITOR'S CONCERN:
\appendix

\section{Bayesian estimation of the submillimetre flux density
distribution}

The so-called Bayesian interpretation of probability (that probabilities
are a {\it subjective} measure of the plausibility of uncertain
states of affairs)
contrasts with the traditional, Frequentist interpretation
(probabilities describe the relative frequencies of the outcomes of
particular experiments).
By adopting this point of view, and by using Bayes' Theorem
of conditional probabilities, one can develop a powerful method
of deductive reasoning that tempers the idealised 
binary (``true'' vs. ``false'') certainty of classical logic, 
to a formalism that is better adapted to cope with the uncertainties
inherent in scientific experimentation: a form of
``quantitative epistemology'' (Loredo, 1990).
Bayesian techniques and philosophies are rapidly gaining ground in 
many areas of astrophysics.
The interested reader is referred to the numerous
reviews of the subject: see, in particular, Loredo (1990, 1992),
or an introductory statistical text such as Wall \& Jenkins (2003).

First, we must assume a functional form for the underlying submm flux
distribution. In this paper, we have investigated three common functions,
each characterised by a single pair of parameters, a characteristic
flux $S_0$ and a width or a slope ($\Sigma$, $\Sigma_{10}$ or $\alpha$):
(i) a lognormal distribution
\begin{equation}
n(S) \propto S^{-1} exp(-(\log{S}-\log{S_0})^2/2\Sigma_{10}^2),
\end{equation}

\noindent
(ii) the Schechter function (Schechter, 1976) 
\begin{equation}
n(S) \propto \left(\frac{S}{S_0}\right)^{\alpha}exp(-S/S_0),
\end{equation}

\noindent
and (iii) the Normal (Gaussian) distribution
\begin{equation}
n(S) \propto exp(-(S-S_0)^2/2\Sigma^2),
\end{equation}

\noindent
(The Gaussian distribution is unphysical in the sense that it 
attributes {\it negative} luminosities to submm sources. 
We include it here as a diagnostic of the method.)
All are normalised so to give the correct total number
of observed sources.

Our objective is to estimate the probability distribution--- our 
subjective knowledge--- over the parameter pairs 
($S_0$; $\Sigma$, $\Sigma_{10}$, $\alpha$), by deriving the
observational outcomes of each distribution and comparing with 
each data point in turn.
The Bayesian method provides a way of updating this knowledge in the light
of new data. 
First, for each function separately,
we have to adopt a prior probability distribution over the
parameters--- call this $P(M_j)$, where each model $M$ 
(described by pairs of ($S_0$; $\alpha$, $\Sigma$ or $\Sigma_{10}$) 
is denoted by the index j. 
In this paper, we have adopted uniform prior probability distributions,
but in general one is free to take into account information from
other sources in constructing the prior.
Next (again, separately for each functional form)
we iteratively update the probabilities in the light of each 
successive data point denoted by index i
(having flux density $S_i$ and rms $\sigma_i$) 
by applying Bayes' theorem:

\begin{equation}
P(M_j | S_i,\sigma_i ) = P(M_j)\times
\frac{P(S_i | M_j,\sigma_i )}
{\Sigma_k P( S_i,\sigma_i | M_k )}
\end{equation}
 
\noindent
where $P(S_i | M_j,\sigma_i )$ is the probability that,
assuming model $M_j$ to be true, we would have
measured a flux $S_i$ for the $i$th target--- 
in this case, given by a convolution of the underlying flux
distribution (given by $M_j$) with the error distribution describing the
scatter of each data point (assumed to be Gaussian, with rms $\sigma_i$).
The denominator acts as a normalising factor.
$P(M_j | S_i,\sigma_i)$ now replaces $P(M_j)$ as our estimate of
the probability function over the parameters. We iterate the procedure
with the next data point, using this new estimate of $P(M_j)$ on the
right hand side, and repeat until all data points have been taken into account.
Note that for this purpose, all data are taken at face value, regardless 
of significance: their measurement error is taken into account
in estimating $P(S_i | M_j,\sigma_i )$,
alleviating the need to draw a sharp distinction between
detection and non-detection.

If we are interested in the value of one parameter in particular---
such as the characteristic flux, $S_0$--- we can use the technique
of {\it Bayesian marginalization} to integrate the final joint
probability distribution ($P(S_0; \alpha,\Sigma_{10},\Sigma)$) with respect to 
the remaining parameters (dubbed ``nuisance parameters'' 
in Bayesian analysis). The resulting marginal distribution ($P(S_0)$)
is the probability density function describing our subjective knowledge 
concerning the true value of the parameter ($S_0$).
(N.B., this marginal distribution--- e.g. top right panel of Figure 4---
should {\it not} be confused with the ``best guess'' 
flux distribution (e.g. bottom panels of Figure 4) obtained by 
inserting the most likely values of the parameters into the function.
The former is a measure of our subjective knowledge (or ignorance), 
the latter our best estimate of the real, objective distribution 
exhibited by the sample.)

The strength of the Bayesian approach is that we can take into account
{\it all} the data, without making any 
distinction between ``detection'' and ``non-detection''. This is particularly
useful for treating low signal-to-noise data,
for combining datasets which have a wide range of errors, or non-Gaussian
errors, or which suffer from different observational biases (such bias
could in principle be taken into account in estimating 
$P(S_i | M_j,\sigma_i )$ for each data point). 
Its weakness is its reliance upon our making an assumption about the
functional form of the underlying distribution
(The final $P(M)$ is, strictly, a {\it conditional} probability, 
the probability
of the parameters taking particular values assuming the underlying form
of the distribution to be true).
This assumption may, in general, be motivated by a particular physical model,
if independent evidence makes the model {\it a priori} more likely.
For the present purposes, it is sufficient that the fit provide an adequate
phenomological description of the data, in order that the key quantities
can be estimated.

\end{document}